**Light nuclei production in heavy ion collisions**

*K. H. Khan[1], M. K. Suleymanov[1,2], Z. Wazir[1], E. U. Khan[1], Mahnaz Q. Haseeb[1], M. Ajaz[1]*

[1]CIIT, Islamabad (Pakistan), [2]JINR, Dubna (Russia)

## 1. Introduction

The creation of new phases of strongly interacting matter is the most interesting area of research for physicist for last many years. Physicists are interested in studying characteristics of newly formatting matter under extreme conditions. The one way to create these new phase is the heavy ion collision at relativistic and ultra relativistic energies. We are interested in centrally dependence of hadrons – nucleus and nucleus -nucleus collisions. These experiments indicate the regime change at some values of the centrality as some critical phenomena. If the regime change Observed in the different experiments takes place unambiguously twice, this would be the most direct experimental evidence to a phase transition from hadronic matter to a phase of deconfined quarks and gluons. After point of regime change the saturation is observed. The simple models cannot explain the effect. For this it is necessary to suggest that the dynamics is same for all such interactions independent of energy and mass of the colliding nuclei and their types. The mechanism to describe the phenomena may be statistical or percolative due to their critical character[1]. The effect depend weakly on the mass of the colliding nuclei so the belief that the mechanism to explain the phenomena may be the percolation cluster formation[2-4]. There is a very great chance that the effect of the light nuclei emission[5-8] in heavy ion collisions may of the accompanying effects of percolation cluster formation and decay. So this effect may be use as some signal on percolation cluster formation.

## 2. Light nuclei production.

There are two types of light nuclei emitted in heavy ion collisions: first type is light nuclei which were produced as a result of nucleus disintegration of the colliding nuclei; second one is light nuclei which were made of protons and neutrons (for example as a result of coalescence mechanism) which were produced in heavy ion interaction at freeze out state. Here we describe the two type of collision to understand the coalescence mechanism.
Peripheral Collision.

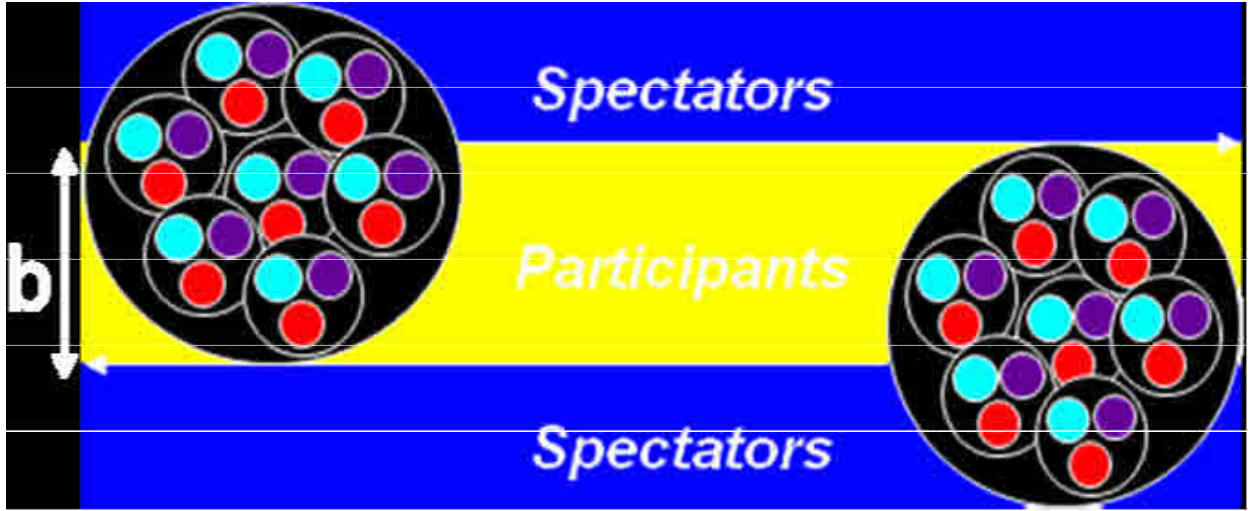

Fig.1

In peripheral collision (Fig.1) nuclei collide in such a way that their some nucleon interact each other known as participants, the nuclei which do not interact are known as spectators and make separate fragments. These fragments may be light or intermediate nuclei. This process is also called fragmentation in Central Collision.

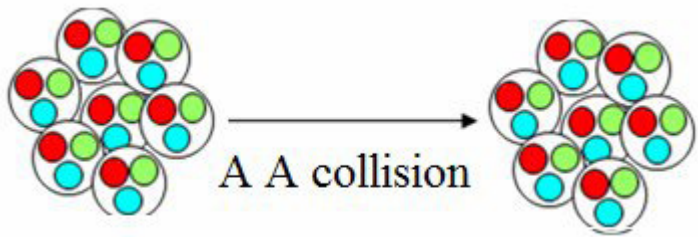

Fig.2

In second type nuclei collide centrally (see Fig.2) in such a way that their maximum nucleons interact with each other and colliding nuclei waste their individuality. So the fragmentation has to suppress in these collisions. It is expected that mainly in these collisions nuclear mater could change into Quark Gluon Plasma (QGP) as a result of increasing temperature and density of nuclear matter.

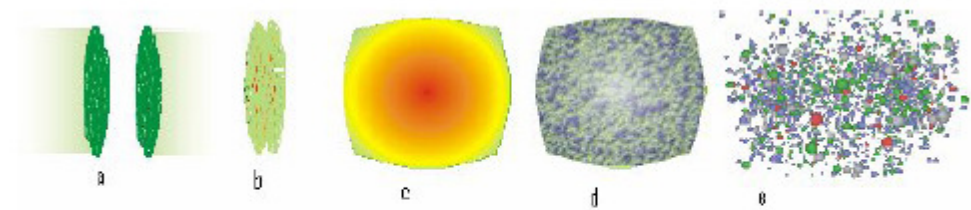

Fig.3

We can see from above figures that as increasing the centrality fragmentation decreases. To understand the different stages of nuclear collision and phase transition mechanism we use the following figure of central collisions. We are working at last stages (e) in this stage we get both type of light nuclei may be formed as a result of coalescence effect due to high pressure as cluster formation and percolation effect and as a result of fragmentation.

We can distinguish the two types of light nuclei by time of detection because fragments are detected earlier than nuclei formed as a result of coalescence. We can confirm Coalescence mechanism by measuring transparency effect for light nuclei at stages (d), if transparency at stage (d) for light nuclei is minimum and we measure light nuclei at stage (e) which is direct confirmation of coalescence effect. Some models are used to measure the light nuclei as a result of coalescence is described bellow;

## 3. Models

### 3.1. Simple Coalescence model

In 1963, Butler and Pearson developed a model for deuteron formation in proton-nucleus Collisions according to them[9]. The key result is that, on account of simple momentum phase space considerations, the deuteron density in momentum space, $d^3N_d/dP^3_p$, is proportional to the proton density in momentum space, $d^3N_p/dp^3_p$ times the neutron density in momentum space, $d^3N_n/dp^3_n$, at equal momentum per nucleon (P=2p) and can be expressed as:

$\gamma d^3N_d/dP^3_d = B_2 (\gamma d^3N_p/dp^3_p)(\gamma d^3N_n/dp^3_n)$

Since many experiments measure protons but not neutrons, it is useful to rewrite this equation assuming the neutron and proton densities to be identical:

$\gamma d^3N_d/dP^3_d = B_2 (\gamma d^3N_p/dp^3_p)^2$

Where

$B_2 = |V0|^2 p (1+m^2/p^2) j(pR)$

However, the underlying phase space relationship survives, and is expressed in a form generalized for nuclear species as:

$\gamma d^3N_A/dp^3_d = B_A (\gamma d^3Np/dp^3_p)^A$

Where

$B_A = (2S_A+1/2^A) 1/N! 1/Z! (R_{np})^N (4\pi p_0^3/3)^{A-1}$

SA is the spin of the cluster of mass A, and N and Z is the neutron and proton numbers of the composite particle. The factor Rnp is the ratio of neutrons to protons participating in the collision.

### 3.2. Density Matrix Model.

The previous models were improved by accounting for some of quantum mechanical aspects of coalescence [10] the particles distributions are represented by density matrices and the wave functions of the particles are assumed to be Gaussian in shape. By considering the internal coalescence in momentum space, the coalescence parameter for fragment of a mass A is

$$B_A = \frac{2s_A+1}{2^A}(R_{np})^N A^{3/2}(4\pi \frac{\nu_A \nu}{\nu_A + \nu})^{3/2(A-1)}$$

### 3.3. Thermodynamic Model.

A thermodynamic model of heavy ion collisions was developed to describe the production of light nuclei in a rapidly expanding system of nucleons [11, 12]. In this model the resulting invariant multiplicity of the composite particles can be expressed as:

$d^3N_A/dp^3_d = (2S_A+1)/2^A [(2\pi)^3/V]^{A-1} e^{-E0/T} (R_{np})^N (d^3N_p/dp^3_p)^A$.

There are many other models but we shall not discuss here.

## 4. Experimental results

Here we discuss experimental results in comparison with model Fast MC results. The results presented in [13] on the measured yield of proton and light nuclei. The formation of light nuclei near central rapidity in a nucleus-nucleus collision requires the constituent nucleons to be close together in phase space in order to be able to coalesce.

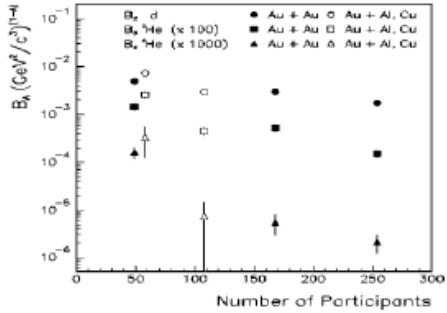

Fig.4 *Dependence of the $B_A$ parameter on the number of participants in Au+Al, Au+Cu, and Au+Au reactions*

The relative probability for the formation of a nuclear cluster is quantified in the BA parameter, this is defined as the proportionality constant between the yields of a nucleus of mass number A and the $A^{th}$ power of the proton yield:

$$E_A \frac{d^3 N_A}{d\vec{p}_A^3} = B_A (E_p \frac{d^3 N_p}{d\vec{p}_p^3})^A$$

Figure below shows the quantity $B_A$ for various targets and event geometries in Au+Au reactions. The data points correspond to events in different centrality bins, which are plotted against the mean number of participating nucleons for that bin as determined by Monte Carlo simulation. In data set, $B_2$ decreases by a factor of 2 from the most peripheral to the most-central bin, while $B_3$ decreases by a factor of 10 and $B_4$ by a factor of 100

## 5. Fast Mc results and Discussion.

Here we have some results obtain by Fast Hadron Freeze-out Generator Model [14]. We consider the dependence of nuclear modification factor on three variables Thermal Freeze out temperature, Chemical freeze out and baryon potential. We defined nuclear modification factor as the ratio of baryons multiplicity at central to peripheral collision. These are some results obtained by simulation With Fast Hadron Freeze–out Generator.

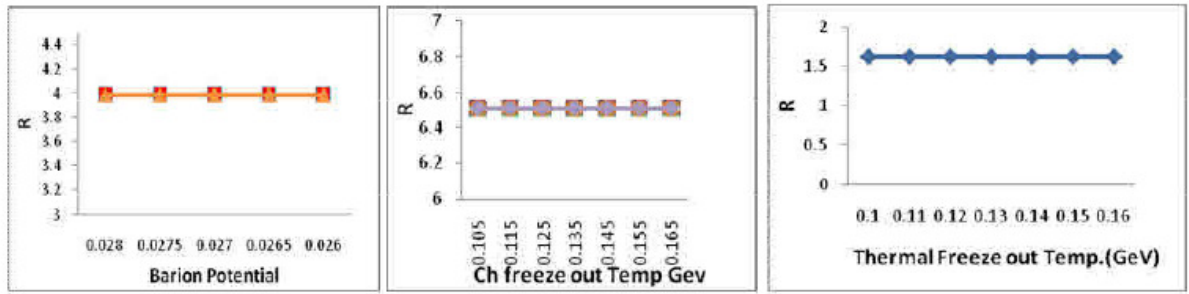

Fig.5 *Dependences of nuclear modification factor on different variables*

In fig.5 R nuclear modification Factor has no dependence on the mentioned variables. FAST MC model does not include nuclear coalescence effect so we shall work in other way. Now we simulate similar data with a model which includes nuclear coalescence effect and then we shall compare these with experimental results. We expect some deviation in above variables in experimental results as compare to simulated one. On this deviation we shall try to explain nuclear coalescence effect. However these variables may have no dependence on nuclear coalescence effect then we shall consider some other variables in same way and explain nuclear coalescence effect.

## 6. References.

[1] Höhne, Claudia. System-size dependence of strangeness production, canonical strangeness suppression and percolation of strings. GSI Darmstadt, Probing QCD with High Energy Nuclear Collisions, Hirschegg 2005.
[2] Satz, H.Parton Percolation in Nuclear Collisions. arXiv: hep-ph/0212046
[3] Brzychczyk, Janusz. Order parameter fluctuations in percolation: Application to nuclear multifragmentation. arXiv: nucl-th/0407008
[4] Pajares, C. String and Parton Percolation. arXiv: hep-ph/0501125.
[5] Adler, S.S., et al. Deuteron and antideuteron production in Au+Au collisions at sqrt (s_NN) =200 GeV. arXiv: nucl-ex/0406004.
[6] Zhangbu Xu for the E864 Collaboration. Production of Light (anti-)Nuclei with E864 at the AGS. arXiv: nucl-ex/9909012.
[7] Albergo, S., et al.2002. *Light nuclei production in heavy-ion collisions at relativistic energies.*Phys. Rev. C, v. 65, 034907.
[8] Bearden, I.G., et al. Deuteron and triton production with high energy sulphur and lead beams. European Physical Journal C; EPJC011124.
[9] Butler, S.T., and Pearson, C.A., 1963. Deuterons from High-Energy Proton Bombardment of Matter. Phys. Rev 129, 836-842
[10] Sato, H., and Yasaki, K., 1981.Cluster formation with 4pi detectors. phys.Letter.B 98.153.
[11] Mekjian, A.Z., 1978. Explosive nucleosynthesis, equilibrium thermodynamics, and relativistic heavy-ion collisions. Phys. Rev. C 17, 1051-1070.
[12] Gupta, S. Das, and Mekjian, A.Z., 1981. Phys. Rep. 72, 131.
[13] Anticic, T.,et al.2000 .Measurements of light nuclei production in 11.5*A* GeV/*c* Au+Pb heavy-ion collisions. Phys. Rev.C 61 064908.
[14] Amelin, N.S., 2006 .A Fast Hadron Freeze-out Generator. arXiv: nucl-th/0608057v4